
\overfullrule=0pt  
\texsis
\preprint
\Eurostyletrue
\superrefsfalse

\def\Acknowledgements{{\bigskip\leftline{{\bf Acknowledgments}}\medskip}}
\referencelist

\reference{Duganrandall}
M.~J.~Dugan, L.~Randall,
\journal Nucl. Phys. B;382,419(1992)
\endreference

\reference{Zaragoza}
J.L.~Alonso, Ph.~Boucaud, J.L.~Cortes and E.~Rivas,
\journal
Phys. Rev. D;44,3258(1991)
\endreference

\reference{*Zaragoza}
\journal
Mod. Phys. Lett. A;5,275(1990)
\endreference

\reference{*Zaragoza}
\journal Nucl. Phys. B {\rm (Proc. Suppl.)};17,461(1990)
\endreference

\reference{ZaragozaII}
For a comprehensive update see:

J.L.~Alonso, Ph.~Boucaud, J.L.~Cortes, F.~Lesmes and E.~Rivas,

proceedings of the topical workshop ``Non perturbative aspects of
Chiral Gauge Theories'', Accademia Nazionale dei Lincei, Rome, 9-11
March 1992
\endreference

\reference{Takeuchi}
T.~Takeuchi,

in 1991 Nagoya Spring School on Dinamical Symmetry breaking,
SLAC--PUB-5619
\endreference

\reference{*Takeuchi}
M.~E.~Peskin and T.~Takeuchi,
\journal Phys. Rev. D;46,381(1992)
\endreference

\reference{Montvay}
I.~Montvay,
\journal Phys. Lett. B;199,89(1987)
\endreference

\reference{*Montvay}
\journal Phys. Lett. B;205,315(1988)
\endreference

\reference{Smit}
J.~Smit,
\journal Acta Physica Polonica B;17,531(1986)
\endreference

\reference{*Smit}
\journal Nucl. Phys. B;175,307(1980)
\endreference

\reference{*Smit}
P.~D.~V.~Swift,
\journal Phys. Lett. B;145,256(1984)
\endreference

\reference{Maiani}
A.~Borelli, L.~Maiani, G.-C.~Rossi, R.~Sisto and M.~Testa,
\journal Nucl. Phys. B;333,335(1990)
\endreference

\reference{MaianiII}
 L.~Maiani,

proceedings of the topical workshop ``Non perturbative aspects of
Chiral Gauge Theories'', Accademia Nazionale dei Lincei, Rome, 9-11
March 1992
\endreference

\reference{*MaianiII}
 L.~Maiani, G.-C.~Rossi and M.~Testa,
\journal Phys. Lett. B;292,397(1992)
\endreference

\reference{Eichten}
E.~Eichten, J.~Preskill,
\journal Nucl. Phys. B;268,179(1986)
\endreference

\reference{Smitswift}
M.~F.~L.~Golterman, D.~N.~Petcher and J.~Smit,
\journal Nucl. Phys. B ;370,51(1992)
\endreference

\reference{*Smitswift}
M.~F.~L.~Golterman and D.~N.~Petcher,
\journal Phys. Lett. B;247,370(1990)
\endreference

\reference{*Smitswift}
\journal Nucl. Phys. B;359,91(1991)
\endreference

\reference{*Smitswift}
\journal Nucl. Phys. B {\rm (Proc. Suppl.)};20,577(1991)
\endreference

\reference{*Smitswift}
M.~F.~L.~Golterman, D.~N.~Petcher and E.~Rivas,
\journal Nucl. Phys. B;377,405(1992)
\endreference

\reference{*Smitswift}
W.~Bock, A.~K.~De and J.~Smit,
preprint HLRZ-91-81, J\"ulich and refs. therein
\endreference                        

\reference{UsEichten}
M.~F.~L.~Golterman and D.~N.~Petcher,
Nucl. Phys. B {\rm (Proc. Suppl.)} \underbar{26} (1992) 486
\endreference

\reference{*UsEichten}
M.~F.~L.~Golterman, D.~N.~Petcher and E.~Rivas,
preprint Wash.~U.~HEP/92-80, to be published in Nucl. Phys.~B
\endreference

\reference{Donadam}
D.~N.~Petcher,
to be published in the proceedings of the 1992 Conference on Lattice
Field Theory, Amsterdam, The Netherlands, and refs. therein
\endreference

\reference{usnonpert}
J.~L.~Alonso, Ph. Boucaud, D.~Espriu, F.~Lesmes and E.~Rivas,
in preparation
\endreference
\endreferencelist

\def\psihat{\widehat{\psi}}
\def\psibar{\overline\psi}
\def\psihatbar{\widehat{\psibar}}

\pubcode{Wash. U. HEP/92-89. \break DFTUZ 92-26}
\titlepage
\title

The $S$, $U$ and $\Delta\rho$  parameters in the Zaragoza proposal for
lattice chiral gauge fermions.
\endtitle

\authors
{\bf Jos\'e Luis Alonso,$^\dagger$ Philippe
Boucaud,$^*$ Jos\'e Luis Cort\'es,$^\dagger$
Felipe Lesmes$^\dagger$ and Elena~Rivas$^\ddagger$}

\bigskip\bigskip
{$\dagger$}{Departamento de F\'{\i}sica Te\'orica,
Universidad de Zaragoza,
50009 Zaragoza, Spain.}

\bigskip
{*}{Laboratoire de Physique Th\'eorique et Hautes Enegies,
Universit\'e Paris XI,
91405 Orsay Cedex, France.}

\bigskip
{$\ddagger$}{Department of Physics,
Washington University,
St. Louis, MO 63130-4899, USA.}
\endauthors
\abstract

Using the Zaragoza proposal for lattice chiral gauge fermions, the
$S$, $U$ and $\Delta\rho$  parameters have been calculated at one
loop. It
is shown that the continuum values for these quantities can be
reproduced without requiring explicit fine tuning of counterterms.
Furthermore, fermion fields doubling is not necessary. To the best of
our knowledge, the Zaragoza proposal is the only scheme which has
these properties.

A necessary (although not sufficient) symmetry is found to support
the calculations. Previous results for some of these parameters in
other lattice chiral regularizations are revisited in the light of
this symmetry.

\endabstract
\bigskip
\vfootnote{$\ddagger$}{address after January $ 1^{th}$, 1993:
Departamento de F\'{\i}sica Te\'orica, Universidad de Zaragoza, 50009
Zaragoza, Spain.}
\endtitlepage


\section{Introduction}

A recent publication by M. Dugan and L. Randall\cite{Duganrandall}
exhibited calculations of the one loop perturbative contribution
to the $S$ parameter in several proposals for implementing lattice
chiral fermions.  In this letter we would like to report the results
for $S$, $U$, and $\Delta\rho$ obtained in the
Zaragoza proposal\cite{Zaragoza}\cite{ZaragozaII}.  This model is
also a candidate for the  description of lattice chiral fermions,
with one major differentiating characteristic with respect to those
studied in \Ref{Duganrandall}: doubler fermions are free and massless
instead of coupled but very massive. This is to be contrasted to
the situation that arises when Wilson terms are used.

\bigskip
In euclidean space the parameters $S$, $U$ and $\Delta\rho$ are
defined by

$$\EQNalign{
\alpha S&\equiv 4e^2
\lbrack\Pi^\prime_{33}(0)-\Pi^\prime_{3Q}(0)\rbrack,\cr
\alpha U&\equiv 4e^2
\lbrack\Pi^\prime_{11}(0)-\Pi^\prime_{33}(0)\rbrack,\EQN{defS}\cr
\Delta\rho&\equiv {-e^2\over s^2 m^2_W}
\lbrack\Pi_{11}(0)-\Pi_{33}(0)\rbrack.\cr}
$$
\bigskip
\noindent
(Here $<W_{a,\mu}(p)W_{b,\nu}(-p)> \equiv \delta_{\mu\nu} \Pi_{ab}(p^2)
+ (p_\mu p_\nu-terms)$; $a,b=1,2,3,Q$ and $\Pi^\prime(0)=d\Pi/dp^2$
at $p^2=0$). These parameters are of great importance because they
describe radiative corrections to the electroweak interaction
observables coming from the Higgs sector\cite{Takeuchi}.

\bigskip
The nice thing about the $S$, $U$, and $\Delta\rho$ parameters is
that they are finite quantities that can be calculated in the
``target'' (unregularized) theory we are aiming to construct, {\it
i.e.} a chiral gauge theory, in particular the Standard Model. The
target continuum Lagrangian is

$$\EQNalign{
{\cal L}^{target} =\quad &{\cal L}^{target}_{A,\phi}\EQN{target}\cr
\quad+&\psibar_L\gamma^\mu(\partial_\mu-iA^L_\mu)\psi_L
+\psibar_R\gamma^\mu(\partial_\mu-iA^R_\mu)\psi_R\cr
\quad+&\psibar_L\phi y\psi_R
+\psibar_R y\phi^\dagger\psi_L.\cr}
$$

\noindent Here, together with the fermion dependent terms,
one has the lagrangian ${\cal L}^{target}_{A,\phi}$ of the bosonic
sector of the Standard Model.

\bigskip
\noindent
For instance, consider a doublet with masses $m_1$, $m_2$, where
$\Delta m\equiv |m_1-m_2|\ll m_1,m_2\approx m$. Using the target
Lagrangian one obtains the following one loop values for $S$, $U$
and $\Delta\rho$

$$\EQNalign{
   S^{target}&={1\over 6\pi},\cr
 U^{target}&={2\over 15\pi}{(\Delta m)^2\over m^2},\EQN{Scont}\cr
 \Delta\rho^{target}&={\alpha \over 12\pi s^2 m^2_W}
 (\Delta m)^2.\cr}
  $$

\bigskip
However, for a given lattice regularization of this target chiral
theory (if indeed any such non--perturbative regularization exists),
there is no guarantee that the results given by \Eq{Scont} will be
reproduced. Therefore these conditions must be imposed (as a
necessary though not sufficient) constraint on the lattice theory
in order to correctly reproduce the ``target theory'' in
the continuum limit.
\bigskip

In their paper Dugan and Randall investigated the value of the $S$
parameter in \Eq{defS} at one loop in perturbation theory for
several lattice chiral fermion proposals. They discussed: Mirror
fermions\cite{Montvay}, the Smit--Swift model\cite{Smit}, and the
Borelli {\it et al.} model\cite{Maiani}.  They concluded that only
the last of these proposals correctly reproduces the continuum
result for $S$ without need of explicit fine tuning of counterterms.

\bigskip
Here, we will see that in the Zaragoza proposal no explicit
fine tuning of counterterms is required in
order to reproduce the continuum one
loop values for the $S$, $U$ and $\Delta\rho$  parameters.  This
result pairs up our model with the Borelli {\it et al.} proposal
studied in \Ref{Duganrandall}. However, in that case, they were
forced to double the number of variables in the regularization
process.

\bigskip
We will identify a symmetry of the target model that appears as a
necessary (though not sufficient) condition to obtain the results
in \Eq{Scont}.  Among the models with Wilson terms
[\use{Ref.Montvay}--\use{Ref.MaianiII}] either the target symmetry
is broken, and so substractions are required, or the symmetry is
preserved by adding extra variables.  (In the last case whether or
not the regularization gives the target values depends on how many
fermions are coupled.)

\noindent
The Zaragoza proposal, because it does not use Wilson terms,
implements this target model symmetry with minimal fermion field
content.
It also keeps the doublers uncoupled,
therefore it gives the target values for $S$, $U$ and
$\Delta\rho$ in perturbation theory.
Explicit one loop calculations will be exhibited.

\section{The Model}

The Zaragoza proposal\cite{Zaragoza}\cite{ZaragozaII} belongs to
a category of models wherein the chiral gauge symmetry is explicitly
broken by the regularization. Therefore additional counterterms
must be included in order to restore gauge invariance in the
continuum limit.

The lagrangian for the standard model is, in euclidean space,

$$
{\cal L}={\cal L}_{U\phi}+{\cal L}_{\psi}+{\cal L}_{\psi_{int}}+
{\cal L}_{ghost}+{\cal L}_{fix}+{\cal L}_{count}.\EQN{lagr}
$$

Where

$$\EQNalign{
{\cal L}_{U\phi}=&
-{1\over g^2}\sum_{\mu\nu} Tr\lbrack U_{\mu\nu}(x)
                                  + h.c.\rbrack\EQN{lagu}\cr
&+ {1\over 2}Tr\lbrack\phi^\dagger(x)\phi(x)
                   +\lambda(\phi^\dagger(x)\phi(x)-1)^2\rbrack\cr
&-{\kappa\over 2}\sum_\mu Tr\lbrack
     \phi^\dagger(x+\hat\mu)U^{L\dagger}_{\mu}(x)\phi(x) U^R_\mu(x)
         +h.c.\rbrack,\cr}
$$
stands for the usual lattice terms corresponding to gauge fields
and boson fields.  Here $\phi(x)$ is the Standard Model Higgs field
in a $2\times 2$ matrix notation.

\bigskip
The lattice action for free fermions is
 $$
 {\cal L}_{\psi}={1\over 2}\sum_{\psi,\mu}(\pm)
\psibar(x)\gamma_\mu\psi(x\pm\hat\mu).\EQN{lagf}
 $$
The $\pm$ signs are correlated and the Hermitian $\gamma$ matrices
satisfy $\lbrace\gamma_\mu,\gamma_\nu\rbrace=2\delta_{\mu\nu}$.
The  Dirac fermion  $\psi$ is in the fundamental representation of
$SU(2)_L\otimes SU(2)_R$

$$
\psi_L={\psi_L^1\choose\psi_L^2}\qquad
\psi_R={\psi_R^1\choose\psi_R^2}.
\EQN{doublet}
$$
The sum  $\sum_\psi$ runs over all $3$ families of leptons and $3
N_c$ families of quarks.

\bigskip
It is important to mention that in the interactions between the
fermions and the gauge and boson fields, described by the term
${\cal L}_{\psi_{int}}$, we wish to arrange things such way that
only the physical fermion is coupled.  This is accomplished by
using in ${\cal L}_{\psi_{int}}$ only the quasi-local field component
$$
\psihat(x)={1\over 2^4}\sum_{y\epsilon h(x)}\psi(y).\EQN{psihat}
$$
Here $h(x)$ is the elemental hypercube that starts from ``$x$'' in
the positive direction.

$$\EQNalign{
{\cal L}_{\psi_{int}}=&
{1\over 2}\sum_{\psi,\mu}
\psihatbar(x)\gamma^\mu
(\Gamma_\mu+{\tilde \Gamma}_\mu)\psihat(x)\EQN{lagfint}\cr
&+ \sum_\psi(\psihatbar_L(x)\phi(x) y_\psi\psihat_R(x)+
               \psihatbar_R(x)y_\psi\phi^\dagger(x)\psihat_L(x)),\cr}
$$
in which
$$\EQNalign{
\Gamma_\mu\psi(x)&
      =\lbrack U_\mu(x)-1\rbrack\psi(x+\hat\mu)-
      \psi(x),\EQN{covariant}\cr
{\tilde \Gamma_\mu}\psi(x)&=
       \psi(x)-\lbrack U^\dagger_\mu(x-\hat\mu)-
       1\rbrack\psi(x-\hat\mu),\cr}
$$
with
$$
U_\mu(x)=U^L_\mu(x)P_L+U^R_\mu(x)P_R,\EQN{U}
$$
$$
U^L_\mu(x)=exp(-iA^L_\mu(x)),\quad U^R_\mu(x)=exp(-iA^R_\mu(x)).
\EQN{ULR}
$$

\noindent
The gauge fields associated with isospin $W^i_\mu$ and hypercharge
$B_\mu$ are introduced as
$$\EQNalign{
A^L_\mu(x)&=
{g\over 2}\sum_{i}\sigma^iW^i_\mu(x)+ {g^\prime\over 2} B_\mu(x)Y_{L}
\EQN{gfields}\cr
A^R_\mu(x)&=
{g^\prime\over 2}B_\mu(x)\left(\matrix{
        Y_{1R}&0\cr
        0&Y_{2R}\cr}\right).\cr}
$$

\noindent
Here the $\sigma^i$ are the pauli matrices, $Y$ denotes the
hypercharge of the different fermions and the Yukawa couplings are
$$
y_\psi=\left(\matrix{y_{\psi_1}&0\cr
                     0&y_{\psi_2}\cr}\right).\EQN{yukawa}
$$

\bigskip
Finally, ${\cal L}_{count}$ includes all the counterterms needed
to recover BRST invariance in the continuum limit. The quantum
theory is defined through the path integral
$$
<(...)>=Z^{-1}\int \lbrack dU\vphantom{\psi^\dagger}\rbrack
\lbrack d\phi\vphantom{\psi^\dagger}\rbrack
\lbrack d\psi\vphantom{\psi^\dagger}\rbrack
\lbrack d\psi^\dagger\rbrack
\lbrack dc\vphantom{\psi^\dagger}\rbrack
\lbrack d{\bar c}\vphantom{\psi^\dagger}\rbrack(...)
 e^{-\sum_x {\cal L}},\EQN{pathint}
$$
where $\lbrack dc\vphantom{\psi^\dagger}\rbrack\lbrack d{\bar
c}\vphantom{\psi^\dagger}\rbrack$ stands for the measure of the
ghost fields and Z is defined so that $<1>=1$.

\bigskip
To understand why we must choose $\psihat$ to describe the fermion
interactions in \Eq{lagfint}, we go to momentum space. There
$\psihat$ is given by
$$
\psihat(q)=F(q)\psi(q),\quad F(q)=\prod_\mu f(q_\mu),\quad
f(q_\mu)=\cos(q_\mu/2),\quad q\epsilon(-\pi,\pi]^4.\EQN{Ffactor}
$$
\noindent
Note that $\psihat$ is the original field $\psi$ modulated by a
form factor $f$ which is responsible for the decoupling of the
doublers at tree level.  This is achieved by forcing the form factor
to vanish for momenta corresponding to the doublers (note that
$f(0)=1$ and $f(\pi)=0$).  In particular \Eq{Ffactor} corresponds
to the most local possible choice.

\bigskip
In summary, this regularization is Hermitian, invariant under
discrete translations and rotations, and no redundant variables
are required. To recover BRST invariance in the continuum limit,
suitable counterterms must be added.  The Lagrangian is constructed
such way that doublers are not coupled,
so no Wilson terms are introduced.  The Lagrangian is also
invariant under a shift transformation of the fermion fields. This
shift--symmetry also assures, if a continuum limit exist, the
non-perturbative decoupling of the doublers. (See \Ref{ZaragozaII}
for more details.)

\section{Calculation of $S$, $U$ and $\Delta\rho$ }

Before starting the calculation, we study what can be learned by
judicious use of symmetries.

Consider, again, the target theory ${\cal L}^{target}$ as given in
\Eq{target}.
For the one loop calculations at hand, the ${\cal L}_{A\phi}$ terms
are not
relevant. The following remarks apply
only to the remainig fermion--dependent piece
of the target  and other lagrangians under discussion.

 Consider the global transformation
$$\EQNdoublealign{
\psi_L(x)&\longrightarrow g_L\psi_L(x),&\cr
\psi_R(x)&\longrightarrow g_R\psi_L(x),&\cr
U_L(x)&\longrightarrow g_L U_L(x)g^\dagger_L,&
\EQN{transf}\cr
U_R(x)&\longrightarrow g_R U_R(x)g^\dagger_R,
\qquad&\hbox{with}\quad  U_{L,R}=\exp(-iA_{L,R})\cr
V(x)&\longrightarrow g_L V(x) g^\dagger_R,
\qquad&\hbox{with}\quad\phi(x)=v+V(x),\cr}
$$
where $g_L=gh^{Y_L}$ for $g\epsilon SU(2)$,
$h\epsilon U(1)$ and
$$
g_R=\left(\matrix{h^{Y_{1R}}&0\cr
        0&h^{Y_{2R}}\cr}\right).\EQN{gR}
$$
(Note that this transformation is not, in general, a global chiral
gauge transformation because here we do not transform the vacuum.)

\bigskip
\noindent
Now, set to zero the Yukawa couplings in the target Lagrangian,
$y\to0$.  At zero Yukawa coupling, ${\cal L}^{target}_{y=0}$ is
invariant under the transformation given in \Eq{transf} in both:

\medskip

\item{(a)} the symmetric phase ($v=0$);

\item{(b)} and the broken phase ($v\neq 0$).

\noindent
One obtains $S=U=\Delta\rho=0$ for either of these two situations.

\bigskip
\noindent
Now, permit  non zero Yukawa couplings in the target model, ${\cal
L}^{target}_{y\neq 0}$.

\medskip

\item{(a)} In the symmetric phase the symmetry \Eq{transf} is
unbroken and the three parameters vanish, $S=\Delta\rho=U=0$.

\item{(b)} In the broken phase, the symmetry \Eq{transf} is broken
and we obtain the results $S^{target}$, $\Delta\rho^{target}$ and
$U^{target}$ of \Eq{Scont}.

\noindent
Therefore, $S$, $U$, and $\Delta\rho$  measure the breaking, by the
Yukawa coupling $y$, of the symmetry of ${\cal L}^{target}_{y=0}$
under the transformation given by \Eq{transf}

\bigskip

Consider now a given regularization for this target model, ${\cal
L}^{reg}$.  The invariance under \Eq{transf} of ${\cal L}^{reg}_{y=0}$
is not {\it a priori} guarantied (not even by gauge invariance).
Therefore:

\medskip

\item{(a)} If ${\cal L}^{reg}_{y=0}$ breaks the symmetry under
\Eq{transf} of ${\cal L}^{target}_{y=0}$, tuning of counterterms will be
required in order to restore it as a necessary condition to get
the target values for $S$, $U$ and $\Delta\rho$ in the continuum
limit.

\item{(b)} When the symmetry is not broken by ${\cal L}^{reg}_{y=0}$,
we need to inspect the situation more carefully.  If the regularization
faithfully translates the Yukawa interaction to the lattice, then
$S$, $U$ and $\Delta\rho$ count the number of fermions present in
the model.

\bigskip

\noindent
In terms of this symmetry, we can analyze all lattice regularizations
for chiral gauge fermions.

\bigskip
For the Zaragoza proposal, the symmetry is unbroken in the regularized
model (before consideration of any counterterms).  Since the doublers
were not coupled, we expect to get the target results for $S$, $U$
and $\Delta\rho$ without tuning of counterterms.

\noindent
Explicit calculations of the one loop contributions to $S$, $U$
and $\Delta\rho$  have been made and they confirm the previous
expectations.

Introduce (``$a$'' being the lattice spacing)

$$\EQNalign{
\Pi^S_{\mu\nu}(ap)&\equiv {4e^2\over \alpha}\lbrack\Pi^{33}_{\mu\nu}(ap)
-\Pi^{3Q}_{\mu\nu}(ap)\rbrack,\EQN{defpis}\cr
\Pi^U_{\mu\nu}(ap)&\equiv {4e^2\over \alpha}\lbrack\Pi^{11}_{\mu\nu}(ap)
-\Pi^{33}_{\mu\nu}(ap)\rbrack.\cr}
$$

\noindent
In terms of these quantities, one has

$$\EQNalign{
\lim_{a\rightarrow 0}
\lbrace
       \half\sum_{\alpha\beta}\left.
{\partial^2\Pi^S_{\mu\nu}(ap)\over \partial ap_\alpha
\partial ap_\beta}\right|_{p=0}
 ap_\alpha ap_\beta\rbrace=&
 (Sp^2+S^\prime p_\mu^2)\delta_{\mu\nu}+K^S  p_\mu p_\nu
,\EQN{SSintegrals}\cr
\lim_{a\rightarrow 0}
\lbrace
       \half\sum_{\alpha\beta}\left.
{\partial^2\Pi^U_{\mu\nu}(ap)\over \partial ap_\alpha
\partial ap_\beta}\right|_{p=0}
 ap_\alpha ap_\beta\rbrace=&
(Up^2+U^\prime p_\mu^2)\delta_{\mu\nu}+K^U p_\mu p_\nu
,\cr
{-\alpha \over 4 s^2 m^2_W}\lim_{a\rightarrow 0}\Pi^U_{\mu\nu}(0)=&
\Delta\rho\delta_{\mu\nu}+K^{\Delta\rho}
.\cr}
$$

\noindent
Here $S$, $U$ and $\Delta\rho$  are those defined in \Eq{defS} and
$S^\prime$, $U^\prime$  are extra terms that may appear because
$\Pi^S_{\mu\nu}$ and $\Pi^U_{\mu\nu}$,  for finite $a$, have
non-covariant expressions.
The $K^S$, $K^U$ and $K^{\Delta\rho}$ are not of interest.
They appear because in
\Eq{defpis} we have not yet extracted the $\delta_{\mu\nu}$
components from
$\Pi^S_{\mu\nu}$ and $\Pi^U_{\mu\nu}$.

$S$, $S^\prime$, $\Delta\rho$, $U$, and $U^\prime$ are rather
complicated integrals. Explicit expressions are given in the
appendix.  We can rigorously prove (see the appendix for some
details) that  the calculation of those integrals leads in fact to
the target results {\it i.e.}
$$\EQNalign{
S&={1\over 6\pi},\cr
U&={2\over 15\pi}{(\Delta m)^2\over m^2},\EQN{results}\cr
\Delta\rho&={\alpha \over 12\pi s^2 m^2_W} (\Delta m)^2.\cr}
$$
and $S^\prime=U^\prime=0$.

\section{Comparison with other models and final remarks}

It is clear that a Wilson term is not invariant under the global
transformation \Eq{transf}.  Such is the case in the Smit--Swift
model. Because of the Wilson terms, ${\cal L}_{y=0}$ for the
Smit--Swift model is not invariant under \Eq{transf}.  Therefore,
one should expect some non--conventional results for this
regularization.

\noindent
Indeed this is what happens, as was first reported in \Ref{Duganrandall}
for the $S$ parameter.  All three parameters have, at one loop,
$r$--dependent expressions $S(r)$, $U(r)$, $\Delta\rho(r)$, where
the $r\rightarrow 0$ limit is $16$ times the target contribution.

\noindent
These results reflect the fact that the doublers contribute to
those quantities through the Wilson terms responsible for breaking
the global symmetry \Eq{transf} at $y=0$.  In the $r\rightarrow 0$
limit, the symmetry is recovered but at the price of having 16
fermions equivalent to each other.  For $r\neq 0$ the symmetry is
broken and the result is in general, a complicated function of $r$.
Counterterms would be necessary, in general, to obtain the target
results.  (Note that, in some particular cases the results for the
Smit-Swift model may agree with the target results. For instance,
when $\Delta m=0$, $U(r)=\Delta\rho(r)=0$ as in \Eq{Scont}, this
is just an accident consequence of a custodial symmetry, also
present in the Smit-Swift model.  Also particular values of $r$
could give the target results, but in general counterterms would
have to be tuned.)

\bigskip
There are ways of building up Wilson terms invariant under symmetry
\Eq{transf}. Examples of this are the Mirror fermions model\cite{Montvay},
and the Borelli {\it et al.} model\cite{Maiani}. In both cases this
is achieved at the expenses of doubling the number of fermion fields
and extending the global transformation \Eq{transf} to include the
extra fields.

\noindent
Furthermore, as reported in \Ref{Duganrandall}, in the Mirror
fermion model (because the mirror fermions are coupled) the mirror
fermions contribute to those quantities as much as the physical
fermions do. On the contrary, for the Borelli {\it et al.} model
(since the extra fermions are not coupled and the global symmetry
is present), the continuum result is achieved.

\noindent
A subsidiary model of the Borelli {\it et al.} model has appeared,
in which redundant variables are not used \cite{MaianiII}. It breaks
the global symmetry, therefore, explicit tuning of
counterterms to recover the symmetry
will be required to get the target values
for $S$, $U$ and $\Delta\rho$.

In our model, since no Wilson terms are required, we get the same
result as in the Borelli {\it et al.} model but without need of
redundant fermion fields. This is a big economy if one were to
implement a numerical simulation.

\bigskip
Of course, many more aspects have to be investigated in any model
of chiral fermions (including those previously mentioned) in order
to conclude whether or not they are good candidates to describe
the target theory. In fact, we now know that the Smit--Swift
model\cite{Smit} and other related models\cite{Eichten}, based on
Wilson--Yukawa terms, do not succeed in describing chiral
fermions,\cite{Smitswift}\cite{UsEichten} and the search for a
fully satisfactory latticization is still open.\cite{Donadam}

So far, the Zaragoza proposal is still a viable candidate for
describing chiral fermions on the lattice.
Unfortunately, we have not yet proved conclusively that it
is in all regards satisfactory.
 In the meantime,
the result presented here can be taken as a technical
improvement:  if the Zaragoza proposal proves to be successful in
describing chiral fermions, we have shown that some of the possible
counterterms that might otherwise have been required will not
actually appear.

The study of the non--perturbative behavior of the
$\Delta\rho$ parameter is of great interest with respect to probing
the effects of heavy fermions in theories with
spontaneous symmetry breaking\cite{ZaragozaII}.
In this regard,
the implications of the results presented in this letter
 are interesting:
since the perturbative calculations of $S$, $U$ and $\Delta\rho$
do not require a fine tuning of counterterms, there is
some hope that in the Zaragoza proposal a full non--perturbative
calculation of these parameters might be
 tractable\cite{usnonpert}.


\Acknowledgements

One of us (E.~R.) wishes to thank the
Washington University Department of Physics
for its hospitality, in particular M. Golterman, D. Petcher
and M. Visser for a critical reading of this manuscript.
The authors want to thank D.~Espriu and B.~Grinstein for
very interesting discussions.
This work is partially supported by the CICYT (proyecto
AEN 90-0030).
E.~R. is supported by a Formaci\'on del Personal Investigador
fellowship from the Spanish government. \vfill



\appendix{A}{Appendix}\label{app}

Using the notation
$$\EQNalign{
F(q)&=\prod_\mu\cos(q_\mu/2),\quad q\epsilon(-\pi,\pi]^4,\quad
\int_q=\int^\pi_{-\pi}d^4 q/(2\pi)^4,\EQN{def1}\cr
s_\mu(q)&=\sin(q_\mu),\quad c_\mu(q)=\cos(q_\mu),\quad
s^2(q)=\sum_\mu\sin^2(q_\mu),\EQN{def2}\cr
D(q)&=s^2(q)+a^2 m^2 F^4(q)\quad \hbox{\rm for
}\quad m=y v
\quad\hbox{\rm in the broken phase}.\EQN{def3}\cr}
$$

\noindent
The expressions for \Eq{defpis} are ($m_1\approx m_2\approx m$,
and $\left|m_1-m_2\right|=\Delta m$)

$$\EQNalign{
\Pi^S_{\mu\nu}(ap)&=\delta_{\mu\nu}\int_q
{{\cal F}^S_\mu(q,ap,m)\over D(q+ap)}
+{\cal O}(\Delta m),\EQN{pisint}\cr
\Pi^U_{\mu\nu}(ap)&=\int_q a^2
{{\cal F}^U(q,ap,m,\Delta m)\over D^2(q+ap)}{\cal G}_{\mu\nu}(q,ap)
+{\cal O}(\Delta m^3),\EQN{piuint}\cr}
$$
where,

$$\EQNalign{
{\cal F}^S_\mu(q,ap,m)&=-16\pi m^2
F^4(q)F^4(q+ap){c^2_\mu(q+ap/2)\over D(q)},\EQN{calfsdef}\cr
{\cal F}^U(q,ap,m,\Delta m)&=+32\pi m^2(\Delta m)^2
F^6(q)F^6(q+ap){1\over D^2(q)},\EQN{calfudef}\cr}
$$
and
$$\EQNalign{
{\cal G}_{\mu\nu}(q,ap)=&c_\mu(q+ap/2)c_\nu(q+ap/2)\EQN{Gdef}\cr
&\left[s_\mu(q)s_\nu(q+ap)+s_\mu(q+ap)s_\nu(q)-
\delta_{\mu\nu}s(q)\cdot s(q+ap)\right].\cr}
$$

\noindent
Using all these previous expressions, \Eq{SSintegrals} can be
written as

$$\EQNalign{
Sp^2+S^\prime p_\mu^2&= \sum_{\alpha\beta}(S^1_{\mu,\alpha\beta}+
S^2_{\mu,\alpha\beta})p_\alpha p_\beta,\EQN{defIs}\cr
K^U p_\mu p_\nu+(Up^2+U^\prime p_\mu^2)\delta_{\mu\nu}&=
\sum_{\alpha\beta}(U^1_{\mu\nu,\alpha\beta}+
U^2_{\mu\nu,\alpha\beta})p_\alpha p_\beta,\EQN{defIu}\cr
\Delta\rho&={-\alpha \over 4\pi s^2 m^2_W}\Pi^U_0,\EQN{defIu0}\cr}
$$
in which

$$\EQNalign{
S^1_{\mu,\alpha\beta}=
    \lim_{a\rightarrow 0}\int_q {a^2\over 2}&
    {1\over D(q)}
(
    {\cal F}^S_{\mu,\alpha\beta}(q)
  -{{\cal F}^S_{\mu,\alpha}(q)D_{,\beta}(q)
  + {\cal F}^S_{\mu,\beta}(q)D_{,\alpha}(q)\over D(q)}
)
,\EQN{A}\cr
S^2_{\mu,\alpha\beta}=
    \lim_{a\rightarrow 0}\int_q {a^2\over 2}&
    {{\cal F}^S_\mu|_{p=0}\over D^2(q)}
(
    {2D_{,\alpha}(q)D_{,\beta}(q)\over D(q)}
     -D_{,\alpha\beta}(q)
)
,\cr
U^1_{\mu\nu,\alpha\beta}=
    \lim_{a\rightarrow 0}\int_q {a^4\over 2}&
    {1\over D^2(q)}
\Biggl\lparen
    {\cal F}^U_{,\alpha\beta}(q){\cal G}_{\mu\nu}|_{p=0}
   +{\cal F}^U_{,\alpha}(q){\cal G}_{\mu\nu,\beta}(q)
   +{\cal F}^U_{,\beta}(q){\cal G}_{\mu\nu,\alpha}(q)
\cr
 &-2{\cal G}_{\mu\nu}|_{p=0}{{\cal F}^U_{,\alpha}(q)D_{,\beta}(q)
   +{\cal F}^U_{,\beta}(q)D_{,\alpha}(q)\over D(q)}
\Biggr\rparen
,\cr
U^2_{\mu\nu,\alpha\beta}=
    \lim_{a\rightarrow 0}\int_q {a^4\over 2}&
    {{\cal F}^U|_{p=0}\over D^2(q)}
\Biggl\lparen
     {\cal G}_{\mu\nu,\alpha\beta}(q)
  -2{{\cal G}_{\mu\nu,\alpha}(q)D_{,\beta}(q)
  + {\cal G}_{\mu\nu,\beta}(q)D_{,\alpha}(q)\over D(q)}
\cr
&+2{{\cal G}_{\mu\nu}|_{p=0}\over D(q)}
\left[3{D_{,\alpha}(q)D_{,\beta}(q)\over D(q)}-D_{,\alpha\beta}(q)\right]
\Biggr\rparen
,\cr
\Pi^U_0=
    \lim_{a\rightarrow 0}\int_q a^2&
     32\pi m^2(\Delta m)^2
     F^{12}(q){(2s_1^2(q)-s^2(q))c_1^2(q)\over (D^2(q))^4 }
,\cr}
$$
and the notation is

$$
O_{,\alpha}(q)=\left. {\partial O(q,ap)\over \partial ap_\alpha}
\right|_{p=0}\quad
O_{,\alpha\beta}(q)=\left. {\partial^2 O(q,ap)\over \partial ap_\alpha
\partial ap_\beta}
\right|_{p=0}.\EQN{deriv}
$$

\bigskip
\noindent
The $K^U$ term in \Eq{defIu} is not of interest. It appears because
we have not yet extracted the $\delta_{\mu\nu}$ component from
$\Pi^U_{\mu\nu}$ in \Eq{piuint}.

\bigskip
The calculation of integrals $\Pi^U_0$, $S^{1,2}_{\mu,\alpha\beta}$,
and $U^{1,2}_{\mu\nu,\alpha\beta}$ proceeds as follows:  each
integral can be split into two integrals, one interior to and one
exterior to an $\epsilon$--ball $=\lbrace q; \quad |q|=\epsilon\rbrace$
with $\epsilon\gg a$.

\bigskip
\noindent
Using Lebesgue's theorem of dominated convergence we can show that
in the exterior integrals the limit $a\rightarrow 0$ and the
integration can be exchanged.  As an easy consequence of this we
can prove that all of the exterior integrals give zero contribution.

\noindent
To apply this theorem we have to place a Lebesgue integrable upper
bound on the modulus of the various integrands  in \Eq{A}.  In this
regularization we can use the bound

$$
{F^2(q)\over D(q)}\leq ({\cos^4(\epsilon/2)\over 2\sin\epsilon})^2
\quad \hbox{\rm for}\quad |q|\geq \epsilon,
\EQN{bound}
$$
\bigskip

\noindent to successfully get upper bounds for the integrands in \Eq{A}.

\noindent As an example, for $\Pi^U_0$

$$
\left|F^{12}(q){(2s^2_1(q)-s^2(q))c^2_1(q)\over (D^2(q))^4}\right|\leq
({\cos^4(\epsilon/2)\over 2\sin\epsilon})^6
\quad\hbox{\rm for}\quad |q|\geq \epsilon.
\EQN{Abound}
$$
\bigskip

\noindent
Here the right hand side is a Lebesgue integrable function.  A
quite similar analysis suffices for the other integrals in \Eq{A}.
Thus we have rigorously proved that when calculating exterior
contributions to \Eq{A}, the limit $a\rightarrow 0$ can be taken
prior to do the integration.  Such limit tells us that  all the
exterior contributions to \Eq{A} vanish.

\medskip
The vanishing of the exterior integrals is a consequence of the
global symmetry in \Eq{transf} present in the Zaragoza proposal
when $y=0$.  It can be understood in the following way:  the
Lagrangian ${\cal L}_{y=0}$ (for ${\cal L}$ given in \Eq{lagr}) is
invariant under the global transformation in \Eq{transf}, so
$S=U=\Delta\rho=0$ for $y=0$.

\noindent
Now, for the total Lagrangian we have the one loop expressions
(\Eq{A}) for $S$,$U$ and $\Delta\rho$.  When calculating the exterior
contributions, because the momenta are of order ${\cal O}(1/a)$,
we can neglect all the Yukawa couplings.  Therefore, the exterior
integrals do not break the global symmetry and no contribution  to
$S$, $U$ and $\Delta\rho$ should be expected from them.

\bigskip
Over the inner region, the limit $a\rightarrow 0$ of the integrals
is easy to do. Things have been arranged such that there is no
contribution from integrals $S^1_{\mu,\alpha\beta}$ and
$U^1_{\mu\nu,\alpha\beta}$
inside the $\epsilon-ball$. All the other integrals in \Eq{A} can
be calculated by picking up the first term in a Taylor expansion
around $q\approx 0$.

\noindent
Finally, the results are

$$
S^1_{\mu,\alpha\beta}=0,\quad
U^1_{\mu\nu,\alpha\beta}=0,\EQN{results1}
$$
and introducing $U^2_{\mu\nu,\alpha\beta}\equiv
A_{\mu\nu,\alpha\beta}+
U^2_{\alpha\beta}\delta_{\mu\nu}$,
$$\EQNdoublealign{
S^2_{\mu,\alpha\beta}&=\delta_{\alpha\beta}&{1\over 6\pi},
\EQN{results2}\cr
U^2_{\alpha\beta}&=\delta_{\alpha\beta}&{2\over 15\pi}
{(\Delta m)^2\over m^2},\cr
\Pi^U_0&=-&{1\over 3\pi }(\Delta m)^2,\cr}
$$
so that \Eq{results} is accomplished.

\vfill
\eject
\center{{\bf References}}
\bigskip
\ListReferences

\bye